\documentclass[superscriptaddress,aip,amsmath,amssymb,floatfix,reprint,raggedbottom]{revtex4-1}
\usepackage[pdftex]{graphicx}
\usepackage{amsmath}
\usepackage{lipsum}
\usepackage{color}
\usepackage{soul}
\usepackage{balance}
\usepackage{subfigure}
\usepackage{fancyhdr}
\usepackage{braket}
\usepackage{threeparttable}
\usepackage{booktabs}
\usepackage{enumitem}
\usepackage{listings}

\begin{document}
\title{Hardware Design for Autonomous Bayesian Networks}
\author{Rafatul Faria}
\email{rfaria@purdue.edu}
\author{Jan Kaiser}
\author{Kerem Y. Camsari}
\author{Supriyo Datta}
\email{datta@purdue.edu}
\affiliation{Department of Electrical and Computer Engineering, Purdue University, West Lafayette, IN, 47906 USA}
\date{\today}

\begin{abstract}
Directed acyclic graphs or Bayesian networks that are popular in many AI related sectors for probabilistic inference and causal reasoning can be mapped to probabilistic circuits built out of probabilistic bits (p-bits), analogous to binary stochastic neurons of stochastic artificial neural networks. In order to satisfy standard statistical results, individual p-bits not only need to be updated sequentially, but also in order from the parent to the child nodes, necessitating the use of sequencers in software implementations. In this article, we first use SPICE simulations to show that  an autonomous hardware Bayesian network can operate correctly without any clocks or sequencers, but only if the individual p-bits are appropriately designed. We then present a simple behavioral model of the autonomous hardware illustrating the essential characteristics needed for correct sequencer-free operation. This model is also benchmarked against SPICE simulations and can be used to simulate large scale networks. Our results could be useful in the design of hardware accelerators that use energy efficient building blocks suited for low-level implementations of Bayesian networks. The autonomous massively parallel operation of our proposed stochastic hardware has biological relevance since neural dynamics in brain is also stochastic and autonomous by nature.

\end{abstract}
\maketitle

\section{Introduction}
\label{section:Introduction}

Bayesian networks (BN) or belief nets are probabilistic directed acyclic graphs (DAG) popular for reasoning under uncertainty and probabilistic inference in real world applications such as medical diagnosis \cite{nikovski2000constructing}, genomic data analysis \cite{jansen2003bayesian, friedman2000using, zou2004new}, forecasting \cite{sun2006bayesian, ticknor2013bayesian}, robotics \cite{premebida2017dynamic}, image classification \cite{park2016image, arias2016medical}, neuroscience \cite{bielza2014bayesian} and so on. BNs are composed of probabilistic nodes and edges from \textit{parent} to \textit{child} nodes and are defined in terms of conditional probability tables (CPT) that describe how each \textit{child node} is influenced by its \textit{parent nodes}\cite{pearl2014probabilistic, heckerman1996causal,  koller2009probabilistic, russell2016artificial}. The CPTs can be obtained from expert knowledge and/or machine learned from data \cite{darwiche2009modeling}. Each node and edge in a Bayesian network have meaning representing specific probabilistic events and their conditional dependencies and they are easier to interpret \cite{correa2009comparison} than neural networks where the hidden nodes do not necessarily have meaning. Unlike neural networks where useful information is extracted only at the output nodes for prediction purposes, Bayesian networks are useful for both prediction and inference by looking at not only the output nodes but also other nodes of interest. Computation of different probabilities from a Bayesian network becomes intractable when the network gets deeper and more complicated with child nodes having many parent nodes. This has inspired various hardware implemenations of Bayesian networks for efficient inference \cite{thakur2016bayesian, rish2005adaptive, chakrapani2007probabilistic, jonas2014stochastic,  zermani2015fpga, tylman2016real, weijia2007pcmos, shim2017stochastic, friedman2016bayesian, querlioz2015bioinspired, behin2016building, shim2017stochastic}. In this article we have elucidated the design criteria for an autonomous (clockless) hardware for BN unlike other implementations that typically use clocks.

\noindent Recently a new type of hardware computing framework called Probabilistic Spin Logic (PSL) is proposed \cite{camsari2017stochastic} based on a building block called probabilistic bits (p-bits), that are analogous to Binary Stochastic Neurons (BSN) \cite{ackley1985learning, neal1992connectionist} of the artificial neural network (ANN) literature. p-bits can be interconnected to solve a wide variety of problems such as  optimization \cite{sutton2017intrinsic, borders2019integer}, inference \cite{faria2018implementing}, an enhanced type of Boolean logic that is invertible \cite{ camsari2017stochastic, faria2017low, pervaiz2017hardware, pervaiz2018weighted}, quantum emulation \cite{camsari2018scaled} and in-situ learning from probability distributions \cite{kaiser2020learning}. 

\begin{figure*}[t!]
\centering
\includegraphics[width=0.75\linewidth]{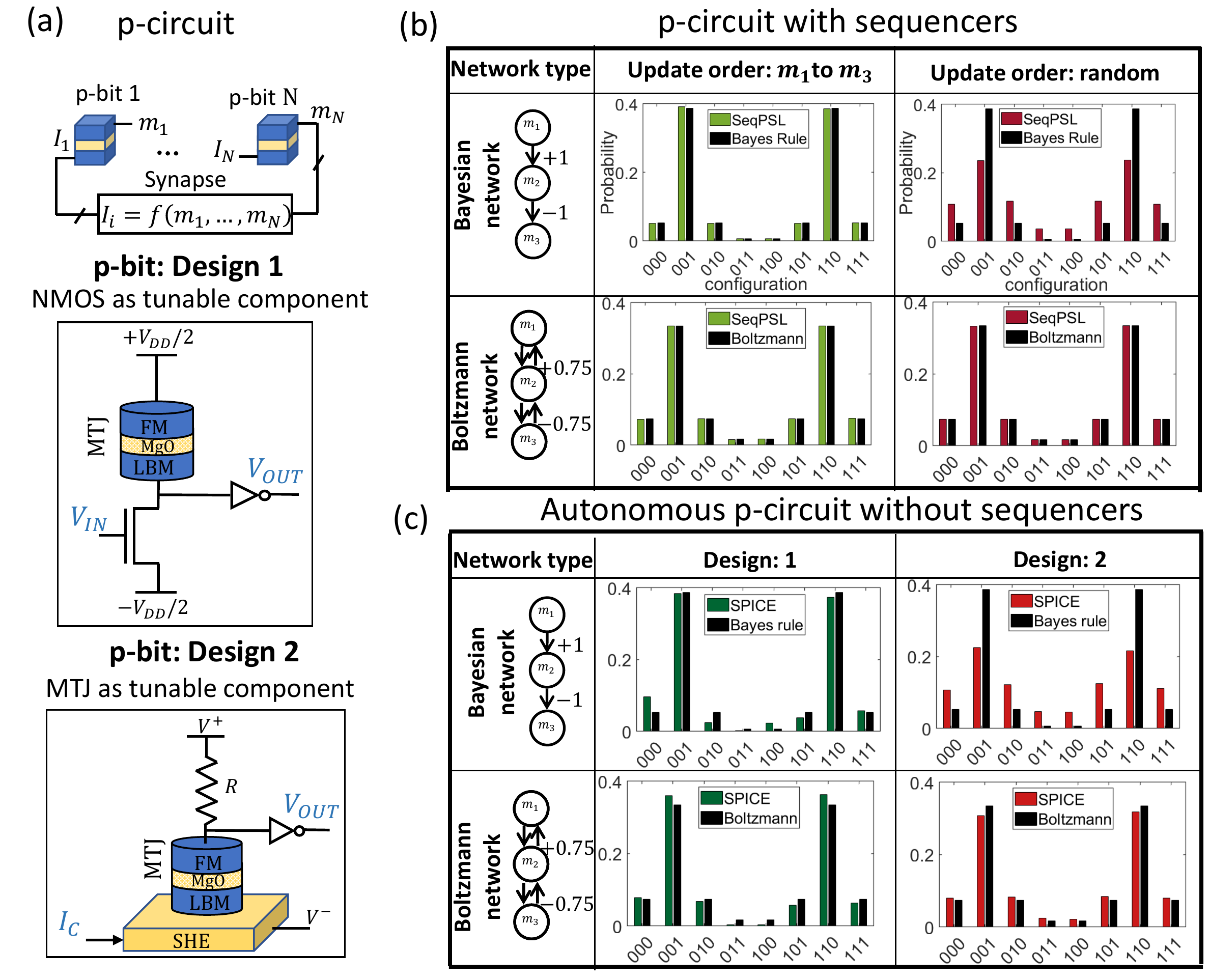}
\caption{\textbf{Clocked versus Autonomous p-circuit}: (a) a probabilistic (p-)circuit is composed of p-bits interconnected by a weight logic (synapse) that computes the input $I_i$ to the $i^{th}$ p-bit as a function of the outputs from other p-bits. Two p-bit designs (design 1 and 2) based on s-MTJ using LBMs have been used to build a p-circuit. (b) Two types of p-circuits are built: a directed or Bayesian network and a symmetrically connected Boltzmann network. The p-circuits are sequential (labeled  as SeqPSL) that means p-bits are updated sequentially, one at a time, using a clock circuitry with a sequencer. It is shown that for Boltzmann networks update order does not matter and any random update order would produce the correct probability distribution. But for Bayesian networks, a specific, parent-to-child update order is necessary to converge to the correct probability distribution dictated by the Bayes rule. (c) The same Bayesian and Boltzmann p-circuits are implemented on an autonomous hardware built with p-bit design 1 and 2 without any clocks or sequencers. It is interesting to note that for Bayesian networks, design 2 fails to match the probabilities from applying Bayes rule, whereas design 1 works quite well as an autonomous Bayesian network.}
\label{fig0}
\end{figure*}

Unlike conventional deterministic networks built out of deterministic, stable bits, stochastic or probabilistic networks composed of p-bits (Fig.~\ref{fig0}a), can be correlated by interconnecting them to construct p-circuits defined by two equations  \cite{ackley1985learning, neal1992connectionist, camsari2017stochastic}: (1) a p-bit/BSN equation and (2) a weight logic/synapse equation. The output of a p-bit, $m_i$ is related to its dimensionless input $I_i$ by the equation:
\begin{subequations}
\begin{equation}
m_i (t+\tau_N) = \mathrm{sgn} \big( \mathrm{rand}(-1,1) + \mathrm{tanh} \ I_i (t) \big) 
\label{eq:CPSLeqn1}
\end{equation}
\noindent where $\mathrm{rand}(-1,+1)$ is a random number uniformly distributed between $-1$ and $+1$, and $\tau_N$ is the neuron evaluation time.

The synapse generates the input $I_i$ from a weighted sum of the states of other p-bits. In general the synapse can be a linear or non-linear function although a common form is the linear synapse described according to the equation:
\begin{equation}
{I_i}(t+\tau_S) = I_{0} \bigg( h_i+\sum_j{J_{ij}m_j(t)} \bigg)
\end{equation}
\label{eq:CPSLeqn2}
\end{subequations}
\noindent where, $h_i$ is the on-site bias and $J_{ij}$ is the weight of the coupling from $j^{th}$ p-bit to $i^{th}$ p-bit and $\tau_S$ is the synpase evaluation time. Several hardware designs of p-bits based on low barrier nanomagnet (LBM) physics have been proposed and also experimentally demonstrated \cite{borders2019integer, ostwal2019spin, ostwal2018spin, debashis2020spintronic, camsari2020charge}. The thermal energy barrier of the LBM is of the order of a few $k_BT$ instead of 40-60 $k_BT$ used in the memory technology to retain stability. Because of thermal noise the magnetization of the LBM keeps fluctuating as a function of time with an average retention time $\tau\sim \tau_0\mathrm{exp}(E_B/k_BT)$ \cite{brown1979thermal}, where $\tau_0$ is a material dependent parameter called attempt time that is experimentally found to be in the range of nanosecond or less and $E_B$ is the thermal energy barrier \cite{lopez2002transition, pufall2004large}. The stochasticity of the LBMs makes them naturally suitable for p-bit implementation. 

Figure \ref{fig0}a shows two p-bit designs: Design 1 (\cite{camsari2017implementing, borders2019integer}) and Design 2  (\cite{camsari2017stochastic, ostwal2019spin}). Design 1 and Design 2 both are fundamental building blocks of STT (Spin Transfer Torque) and SOT (Spin Orbit Torque) MRAM (Magnetoresistive Random Access Memory) technologies respectively \cite{bhatti2017spintronics}. Their technological relevance motivates us to explore their implementations as p-bits. Design 1 is very similar to the commercially available 1T/1MTJ (T: Transistor, MTJ: Magnetic Tunnel Junction) embedded MRAM device where the free layer of the MTJ is replaced by an in-plane magnetic anisotropy (IMA) or perpendicular magnetic anisotropy (PMA) LBM.  Design 2 is similar to the basic building block of SOT-MRAM device \cite{liu2012spin} where the thermal fluctuation of the free layer magnetization of the stochastic MTJ (s-MTJ) \cite{vodenicarevic2017low, vodenicarevic2018circuit, mizrahi2018neural, zink2018telegraphic, borders2019integer, parks2018superparamagnetic} is tuned by a spin current generated in a heavy metal layer underneath the LBM due to SOT effect. The in-plane polarized spin current from the SOT effect in the SHE (Spin Hall Effect) material in design 2 requires an in-plane LBM to tune its magnetization, although a perpendicular LBM with a tilted anisotropy axis is also experimentally shown to work \cite{debashis2020correlated}. Whereas design 2 requires spin current manipulation, design 1 does not rely on that as long as circular in-plane LBMs with continuous valued magnetization states that are hard to pin are used. In-plane LBMs also provide faster fluctuation than perpendicular ones leading to faster sampling speed in the probabilistic hardware \cite{kaiser2019subnanosecond, hassan2019low}.

\begin{figure*}
\centering
\includegraphics[width=0.99\linewidth]{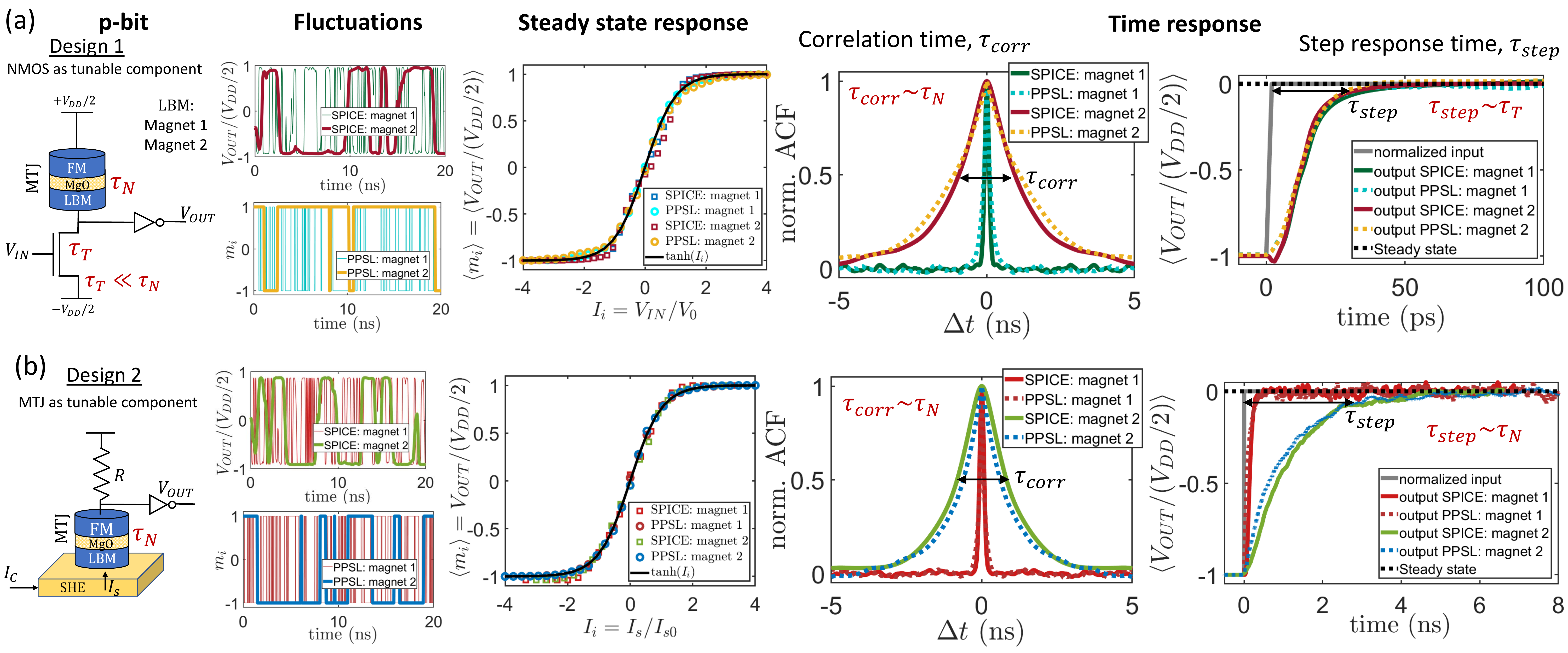}
\caption{\textbf{Autonomous behavioral model for p-bit: Design 1 and 2}: (a) Behavioral model for the autonomous hardware with design 1 is benchmarked with SPICE simulations of the actual device involving experimentally benchmarked modules. The behavioral model (labeled as `PPSL') shows good agreement with SPICE in terms of capturing fluctuation dynamics, steady state sigmoidal response, and two different time responses: autocorrelation time of the fluctuating output under zero input condition labeled as $\tau_{corr}$ which is proportional to the LBM retention time $\tau_N$ in the nanosecond range and the step response time $\tau_{step}$ defined by the transistor response time $\tau_T$ which is few picoseconds and much smaller than $\tau_N$. The magnet parameters used in the simulations are mentioned in section~\ref{section:BehavioralModel} (b) Similar benchmarking for p-bit design 2. In this case $\tau_{step}$ is proportional to $\tau_N$.}
\label{fig1}
\end{figure*}

The key distinguishing feature of the two p-bit designs (design 1 and 2) is the time scales in implementing eqn.~\ref{eq:CPSLeqn1}. From a hardware point of view, eqn.~\ref{eq:CPSLeqn1} has two components: a random number generator (RNG) ($\mathrm{rand}$) and a tunable component ($\mathrm{tanh}$). In design 1, the RNG is the s-MTJ utilizing an LBM and the tunable component is the NMOS transistor, thus having two different time scales in the equation. But in design 2, both the RNG and the tunable component are implemented by a single s-MTJ utilizing an LBM, thus having just one time scale in the equation. This difference in time scales in the two designs is shown in fig.~\ref{fig1}. Note that, although the two p-bit designs have the same RNG source, namely a fluctuating magnetization, it is the difference in their circuit configuration with or without the NMOS transistor in the MTJ branch that results in different time dynamics of the two designs.

 In traditional software implementations, p-bits are updated sequentially for accurate operation such that after each $\tau_S+\tau_N$ time interval, only one p-bit is updated \cite{hinton2007boltzmann}. This naturally implies the use of sequencers to ensure the sequential update of p-bits. The sequencer generates an \textit{Enable} signal for each p-bit in the network and ensures that no two p-bits update simultaneously. The sequencer also makes sure that every p-bit is updated at least once in a time step where each time step corresponds to $N*(\tau_S+\tau_N)$, $N$ being the number of p-bits in the network. \cite{roberts1997updating, pervaiz2018weighted}. For symmetrically connected networks ($J_{ij}=J_{ji}$) such as Boltzmann machines, the update order of p-bits does not matter and any random update order produces the standard probability distribution described by equilibrium Boltzmann law as long as p-bits are updated sequentially. But for directed acyclic networks ($J_{ij}\neq 0, J_{ji}=0$) or Bayesian networks to be consistent with the expected conditional probability distribution, \textit{p-bits need to be updated not only sequentially, but also in a specific update order which is from the parent to child nodes} \cite{neal1992connectionist} similar to the concept of forward sampling in belief networks \cite{henrion1988propagating, guo2002survey, koller2009probabilistic}. As long as this parent to child update order is maintained, the network converges to the correct probability distribution described by probability chain rule or Bayes rule. This effect of update order in a sequential p-circuit is shown on a three p-bit network in fig.~\ref{fig0}b.

Unlike sequential p-circuits in ANN literature, the distinguishing feature of our probabilistic hardware is that it is \textit{autonomous} where each p-bit runs in parallel without any clocks or sequencers. This autonomous p-circuit (ApC) allows massive parallelism potentially providing peta flips per second sampling speed \cite{sutton2019autonomous}. The complete sequencer-free operation of our ``autonomous'' p-circuit is very different from the ``asynchronous'' operation of spiking neural networks \cite{merolla2014million, davies2018loihi}. Although p-bits are fluctuating in parallel in an ApC, it is very unlikely that two p-bits will update at the exact same time since random noise control their dynamics. Therefore persistent parallel updates are extremely unlikely and are not a concern. Note that even if p-bits update sequentially, each update has to be \textit{informed} such that when one p-bit updates it has received the up-to-date input $I_i$ based on the latest states of other p-bits $m_j$ that it is connected to. This informed update can be ensured as long as the synapse response time is much faster than the neuron time ($\tau_S \ll \tau_N$) and this is a key design rule for an ApC. An ApC works properly for a Boltzmann network without any clock since no specific update order is required in this case. But it is not intuitive at all if an ApC would work for a Bayesian network since a particular parent to child \textit{informed} update order is required in this case as shown  in fig.~\ref{fig0}b. As such, it is not straightforward that a clockless autonomous circuit can naturally ensure this specific informed update order. In fig.~\ref{fig0}c, we have shown that it is possible to design hardware p-circuit that can naturally ensure a parent to child informed update order in a Bayesian network without any clocks. In fig.~\ref{fig0}c, two p-bit designs are evaluated for implementing both Boltzmann and Bayesian networks. We have shown that design 1 is suitable for both Boltzmann and Bayesian networks. But design 2 is suitable for Boltzmann networks only and does not work for Bayesian networks in general. The synapse in both types of p-circuits is implemented using a resistive crossbar architecture \cite{alibart2013pattern, camsari2017implementing}, although there are also other types of hardware synapse implementations based on memristors \cite{mansueto2019realizing, li2018efficient, mahmoodi2019versatile}, magnetic tunnel junctions \cite{vaibhav2020synapse}, spin orbit torque driven domain wall motion devices \cite{zand2018low}, phase change memory devices \cite{ambrogio2018equivalent} and so on. In all the simulations $\tau_S$ is assumed to be negligible compared to other time scales in the circuit dynamics.

Our proposed probabilistic hardware for Bayesian networks shows significant biological relevance because of the following reasons: (1) The brain consists of neurons and synapses. The basic building block called ‘p-bit’ of our proposed hardware mimics the neuron and the interconnection among p-bits mimics the synapse function. (2) The components of brain are stochastic or noisy by nature. p-bits mimicking the neural dynamics in our proposed hardware are also stochastic. (3) Brain does not have a single clock for synchronous operation and can perform massively parallel processing  \cite{strukov2019building}. Our autonomous hardware also does not have any global clock or sequencers and each p-bit fluctuates in parallel allowing massively parallel operation.

Further we have provided a behavioral model in section~\ref{section:BehavioralModel} for both design 1 and 2 illustrating the essential characteristics needed for correct sequencer-free operation of BNs. Both models are benchmarked against state-of-the-art device/circuit models (SPICE) of the actual devices and can be used for the efficient simulation of large scale autonomous networks.

\section{Behavioral model for autonomous hardware}

\label{section:BehavioralModel}

\subsection{Autonomous behavioral model: Design 1}
The autonomous circuit behaviour of design 1 can be explained by slightly modifying the two equations (eqns.\ref{eq:CPSLeqn2} a and b) stated in section~\ref{section:Introduction}. The fluctuating resistance of the low barrier nanomagnet based MTJ is represented by a correlated random number $r_{MTJ}$ with values between -1 and +1 and an average dwell time of the fluctuation denoted by $\tau_N$. The NMOS transistor tunable resistance  is denoted by $r_T$ and the inverter is represented by a $sgn$ function. Thus the normalized output $m_i=V_{OUT,i}/(V_{DD}/2)$ of the $i_{th}$ p-bit can be expressed as:
\begin{equation}
m_i\left(t+\Delta t\right)=\mathrm{sgn}\left(r_{T,i}\left(t+\Delta t\right)-r_{MTJ,i}\left(t+\Delta t\right)\right)
\end{equation}
where, $\Delta t$ is the simulation time step, $r_{T,i}$ is the NMOS transistor resistance tunable by the normalized input $I_i=V_{IN,i}/V_0$ where $V_0$ is a fitting parameter which is $\approx 50 \mathrm{mV}$ for the chosen parameters and transistor technology and $r_{MTJ,i}$ is a correlated random number generator with an average retention time of $\tau_N$. $r_{T,i}$ as a function of input $I_i$ is approximated by a $\mathrm{tanh}$ function with a response time denoted by $\tau_T$ modelled by the following equations:

\begin{equation}
\begin{aligned}
r_{T,i}\left(t+\Delta t \right) & = r_{T,i}\left(t\right)\mathrm{exp}\left(-\Delta t/\tau_T\right)+ \\
& \left(1-\mathrm{exp}\left(-\Delta t/\tau_T\right)\right)\left(\mathrm{tanh}\left(I_i(t+\Delta t)\right)\right)
\label{eq:rT}
\end{aligned}
\end{equation}

The synapse delay $\tau_S$ in computing the input $I_i$ can be modelled by:
\begin{equation}
\begin{aligned}
I_i\left(t+\Delta t \right)& =I_i\left(t\right)\mathrm{exp}\left(-\Delta t/\tau_S\right)+\\
& \left(1-\mathrm{exp}\left(-\Delta t/\tau_S\right)\right)\left(I_0\Big(\sum_{j}J_{ij}m_j(t)+h_j\Big)\right)
\end{aligned}
\end{equation}

For calculating $r_{MTJ,i}$ , at time $t+\Delta t$ a new random number will be picked according to the following equations:
\begin{subequations}
\begin{equation}
r_{flip,i}\left(t+\Delta t \right)= \mathrm{sgn}\left(\exp\left(-\frac{\Delta t}{\tau_N}\right)-\mathrm{rand}_{[0,1]} \right)
\label{eq:thetaMTJ}
\end{equation}
where, $\mathrm{rand}_{[0,1]}$ is a uniformly distributed random number between 0 and 1 and $\tau_N$ represents the average retention time of the fluctuating MTJ resistance. If $r_{flip}$ is -1, a new random $r_{MTJ}$ will be chosen between $-1$ and $+1$. Otherwise the previous $r_{MTJ}(t)$ will be kept in the next time step $(t+\Delta t)$, which can be expressed as:

\begin{equation}
\begin{aligned}
r_{MTJ,i}\left(t+\Delta t \right) & =\frac{r_{flip,i}\left(t+\Delta t \right)+1}{2}r_{MTJ,i}\left(t\right) \\
& -\frac{r_{flip,i}\left(t+\Delta t \right)-1}{2}\mathrm{rand}_{[-1,1]}
\end{aligned}
\end{equation}
\end{subequations}

The charge current flowing throught the MTJ branch of p-bit design 1 can get polarized by the fixed layer of the MTJ and generate a spin current $I_{MTJ}$ that can tune/pin $r_{MTJ}$ by modifying $\tau_N$ according to:
\begin{equation}
\tau_N=\tau_N^0 \mathrm{exp}(r_{MTJ}I_{MTJ})
\label{eq:Imtj}
\end{equation}
where, $\tau_N^0$ is the retention time of $r_{MTJ}$ when $I_{MTJ}=0$. This pinning effect by $I_{MTJ}$ is much smaller in in-plane magnets (IMA) than perpendicular magnets (PMA) \cite{hassan2019low}.

Figure.~\ref{fig1}a shows the comparison of this behavioral model for p-bit design 1 with SPICE simulation of the actual hardware in terms of fluctuation dynamics, sigmoidal charateristic response, autocorrelation time ($\tau_{corr}$) and step response time ($\tau_{step}$) and in all cases the behavioral model closely matches SPICE simulationsl. SPICE simulation involves experimentally benchmarked modules for different parts of the device, for example solving stochastic Landau-Lifshitz-Gilbert equation (sLLG) for LBM physics and the 14 nm Predictive Technology Model (PTM) for transistors. The autonomous behavioral model for design 1 is labeled as ``PPSL: design 1''.  The benchmarking is done for two different LBMs: (1) Faster fluctuating magnet 1 with saturation magnetization $M_s=1100$ emu/cc, diameter $D=22$ nm, thickness $th=2$ nm, in-plane easy axis anisotropy $H_k=1$ Oe, damping coefficient $\alpha=0.01$, demagnetization field $H_d=4\pi M_s$ and (2) Slower fluctuating magnet 2 with the same parameters as in magnet 1 except $D=150$ nm. The fast and slow fluctuations of the normalized output $m_i= V_{OUT,i}/(V_{DD}/2)$ are captured by changing the $\tau_N$ parameter in the PPSL model. In the steady state sigmoidal response, $V_0$ is a tanh fitting parameter that defines the width of the sigmoid and lies within the range of $40$ mV to $60$ mV reasonably well depending on which part of the sigmoid needs to be better matched. In fig.~\ref{fig1}, $V_0$ value of $50$ mV is used to fit the sigmoid from SPICE simulation.

There are two types of time responses: (1) Autocorrelation time under zero input condition labeled as $\tau_{corr}$ and (2) step response time $\tau_{step}$. The full width half maximum (FWHM) of the autocorrelation function of the fluctuating output under zero input is defined by $\tau_{corr}$ which is proportional to the retention time $\tau_N$ of the LBM. The step response time $\tau_{step}$ is obtained by taking an average of the p-bit output over many ensembles when the input $I_i$ is stepped from a large negative value to zero at time $t=0$ and measuring the time it takes for the ensemble averaged output to reach its statistically correct value consistent with the new input. $\tau_{step}$ defines how fast the first statistically correct sample can be obtained after the input is changed. For p-bit design 1, $\tau_{step}$ is independent of LBM retention time $\tau_N$ and is defined by the NMOS transistor response time $\tau_T$ which is much faster (few picoseconds) than LBM fluctuation time $\tau_N$. The effect of this two very different time scales in design 1 ($\tau_{step} \ll \tau_{corr}$) on an  autonomous Bayesian network is described in section~\ref{section:Difference}.\\

\subsection{Autonomous behavioral model: Design 2}
The autonomous behavioral model for design 2 is proposed in \cite{sutton2019autonomous}. In this article, we have benchmarked this model with the SPICE simulation of the single p-bit steady state and time responses shown in fig.~\ref{fig1}b. According to this model, the normalized output $m_i=V_{OUT,i}/(V_{DD}/2)$ can be expressed as:
\begin{subequations}
\begin{equation}
m_i(t+\Delta t)=m_i(t)\mathrm{sgn}\Big(p_{NOTflip,i}(t+\Delta t)-\mathrm{rand}_{[0,1]}\Big)
\label{eq:PPSLPRX}
\end{equation}
\begin{equation}
p_{NOTflip,i}(t+\Delta t)=\exp\Big(-\frac{\Delta t}{\tau_N \exp(I_i m_i (t))}\Big)
\end{equation}
\end{subequations}
where, $p_{NOTflip,i})(t+\Delta t)$ is the probability of retention of the $i^{th}$ p-bit (or ``not flipping'') in the next time step that is a function of average neuron flip time $\tau_N$, input $I_i$ and the current p-bit output $m_i(t)$. Figure.~\ref{fig1}b shows how this simple autonomous behavioral model for design 2 matches reasonably well with SPICE simulation of the device in terms of fluctuation dynamics, sigmoidal charateristic response, autocorrelation time ($\tau_{corr}$) and step response time ($\tau_{step}$). In design 2, $\tau_{step}$ and $\tau_{corr}$ are both proportional to LBM fluctuation time $\tau_{N}$ unlike design 1.

Different time scales in p-bit design 1 and 2 are also reported in \cite{hassan2019low} in an energy-delay analysis context. In this article, we explain the effect of these time scales in designing an autonomous Bayesian network (section~\ref{section:Difference}).

\section{Difference between Design 1 and Design 2 in implementing BN}
\label{section:Difference}

\begin{figure}[t!]
\centering
\includegraphics[width=0.99\linewidth]{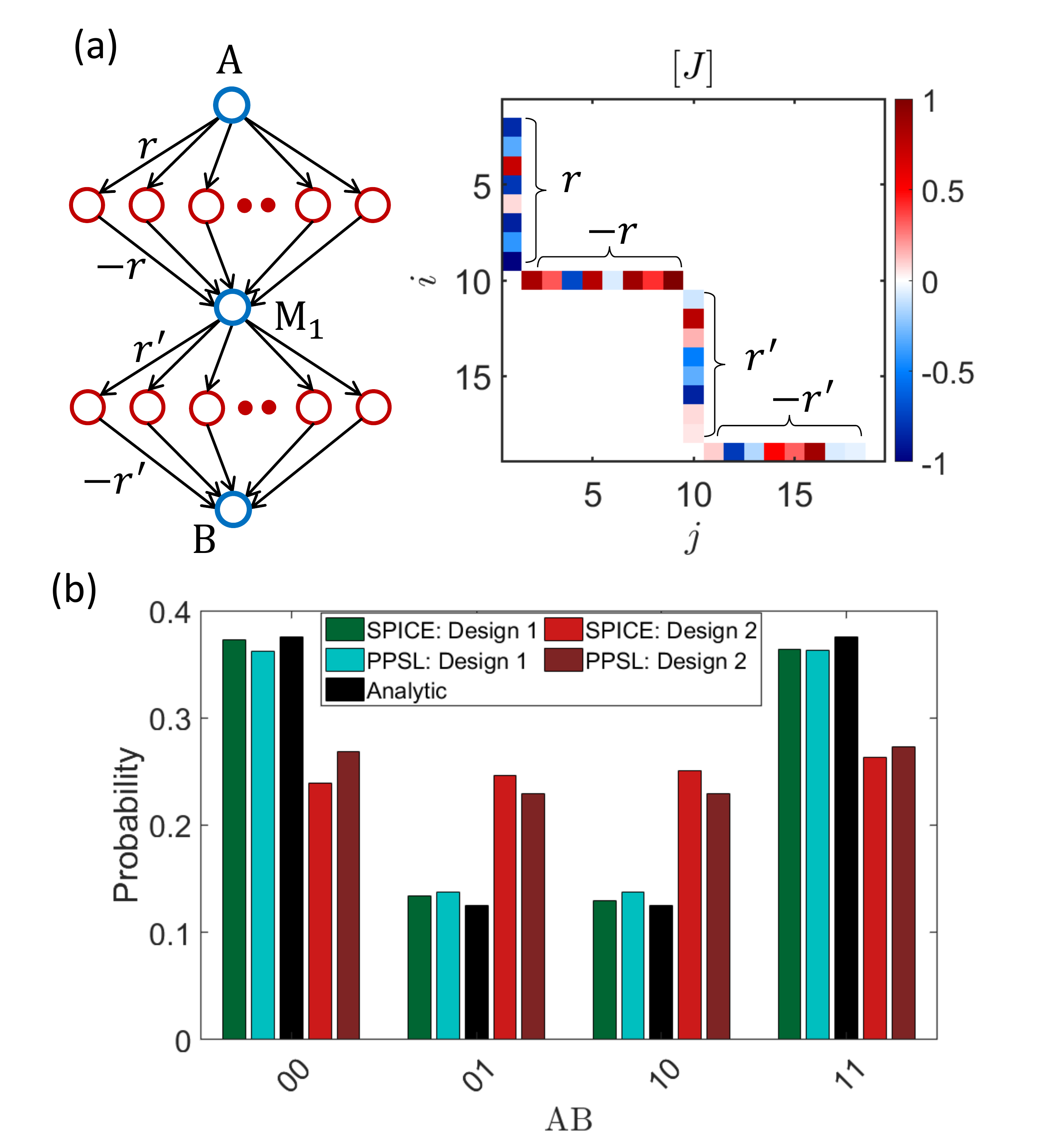}
\caption{\textbf{Difference between design 1 and design 2:} (a) The behavioral models described in fig.~\ref{fig1} are applied to simulate a $19$ p-bit BN with random $J_{ij}$ between +1 and -1. The interconnections are designed in such a way so that pairs of intermediate nodes $(A,M_1)$ and $(M_1,B)$ get anti-correlated and $(A,B)$ gets positively correlated. (b) The probability distribution of four configurations of $AB$ are shown in a histogram from different approaches (SPICE, behavioral model and analytic). The behavioral models for two designs  (labeled as PPSL) match reasonably well with the corresponding results from SPICE simulation of the actual hardware. Note that While design 1 matches with the standard analytical values quite well, design 2 does not works as an autonomous Bayesian network in general.}
\label{fig2}
\end{figure}

The behavioral models introduced in section~\ref{section:BehavioralModel} are applied to implement a multi layer belief/Bayesian network with $19$ p-bits and random interconnection strengths between $+1$ and $-1$ (fig.~\ref{fig2}a). For illustrative purposes, the interconnections are designed in such a way that although there are no meaningful correlations between the blue and red colored nodes with random couplings, pairs of intermediate nodes $(A,M_1)$ and $(M_1,B)$ get negatively correlated because of a net $-r^2$ type coupling through each branch connecting the pairs. So it is expected that the start and end nodes $(A,B)$ get positively correlated. Fig.~\ref{fig2}b shows histograms of four configurations (00, 01, 10, 11) of the pair of nodes $A$ and $B$ obtained from different approaches: Bayes rule (labeled as Analytic), SPICE simulation of design 1 (SPICE: Design 1) and design 2 (SPICE: Design 2), autonomous behavioral model for design 1 (PPSL: Design 1) and design 2 (PPSL: design 2). It is shown that results from SPICE simulation and behavioral model for design 1 matches reasonably well with the standard analytical values showing 00 and 11 states with highest probability whereas design 2 autonomous hardware does not work well in terms of matching with the analytical results and shows approximately all equal peaks. We have tested this basic conclusion for other networks as well with more complex topology as shown in fig.S1 of the supplementary section. The analytical values are obtained from applying the standard joint probability rule for BNs \cite{pearl2014probabilistic, russell2016artificial} which is:
\begin{equation}
P(x_1,x_2,...,x_N)=\prod_{i=1}^{N}x_i|Parents(x_i)
\end{equation}
Joint probability between two specific nodes $x_i$ and $x_j$ can be calculated from the above equation by summing over all configurations of the others nodes in the network which becomes computationally expensive for larger networks. But one major advantage of our probabilistic hardware is that probabilities of specific nodes can be obtained just by looking at the nodes of interest ignoring all other nodes in the system similar to what Feynman stated about a probabilistic computer imitating the probabilistic laws of nature \cite{feynman1999simulating}. Indeed, in the Bayesian network example in fig.~\ref{fig2}, the probabilities of different configurations of nodes $A$ and $B$ were obtained just by looking at the fluctuating outputs of the two nodes ignoring all other nodes. For the SPICE simulation of design 1 hardware, tanh fitting parameter $V_0 = 57$ mV is used and the mapping principle from dimensionless coupling terms $J_{ij}$ to the coupling resistances in the hardware is described in \cite{faria2018implementing}.

The reason why design 1 works for a BN and design 2 does not, is because of the two very different time responses of the two designs shown in fig.~\ref{fig1}. It is this two different time scales in design 1 ($\tau_{step} \ll \tau_{corr}$) that naturally ensures a parent to child informed update order in a Bayesian network. The reason is that when $\tau_{step}$ is small, each child node can immediately respond to any change of its parent nodes that have a much larger time scale $\propto \tau_{corr}$, and thus can be conditionally satisfied with the parent nodes very fast. Otherwise, if $\tau_{corr}$ gets comparable to $\tau_{step}$, the child node will not be able to keep up with the fast changing parent nodes and will produce substantial number of statistically incorrect samples over the entire time range thus deviating from the correct probability distribution.

The effect of $\tau_{step}/\tau_{corr}$ ratio is shown in fig.~\ref{fig3} for the same BN presented in fig.~\ref{fig2}  by plotting the histogram of $AB$ configurations for different $\tau_T/\tau_N$ ratios. It is shown that when $\tau_T/\tau_N$ ratio is small, the histogram converges to the correct distribution. As $\tau_T$ gets comparable to $\tau_N$, the histogram begins to diverge from the correct distribution. Thus the very fast NMOS transistor response in design 1 makes it suitable for an autonomous Bayesian network hardware. One thing to note that under certain conditions, results from design 2 can also match the analytical results if the input $I_i$ to each p-bit in the network always fluctuates between large values that ensures a fast step response time.

So apart from ensuring a fast synapse compared to neuron fluctuation time ($\tau_S \ll \tau_N$) which is the design rule for an autonomous probabilistic hardware, the autonomous Bayesian network demands an additional p-bit design rule which is a much faster step response time of the p-bit compared to its fluctuation time ($\tau_{step} \ll \tau_N$) as ensured in design 1. In all the simulations the LBM was a circular in-plane magnet whose magnetization spans all values between +1 and -1 and negligible pinning effect. If the LBM is a PMA magnet with bipolar fluctutations having just two values +1 and -1, design 1 will not provide any sigmoidal response except with substantial pinning effect \cite{borders2019integer}. Under this condition, $\tau_{step}$ of design 1 will be comparable to $\tau_N$ again and the system will not work as an autonomous Bayesian network in general. Therefore LBM with continuous range fluctuation is expected for design 1 p-bit to work properly as a Bayesian network.

\section{Discussion}
In this article we have elucidated the design criteria for an autonomous clockless hardware for Bayesian networks that requires a specific parent to child update order when implemented on a probabilistic circuit. By performing SPICE simulations of two autonomous probabilistic hardwares built out of p-bits (design 1 and design 2 in fig.~\ref{fig0}), we have shown that the autonomous hardware will naturally ensure a parent to child informed update order without any sequencers if the step response time ($\tau_{step}$) of the p-bit is much smaller than its autocorrelation time ($\tau_{corr}$). This criteria of having two different time scales is met in design 1 as $\tau_{step}$ comes from the NMOS transistor response time $\tau_T$ in this design which is few picoseconds. We have also proposed an autonomous behavioral model for design 1 and benchmarked it against SPICE simulation of the actual hardware. All the simulations using behavioral model for design 1 are performed ignoring some non-ideal effects listed below:

\begin{figure}[t!]
\centering
\includegraphics[width=0.97\linewidth]{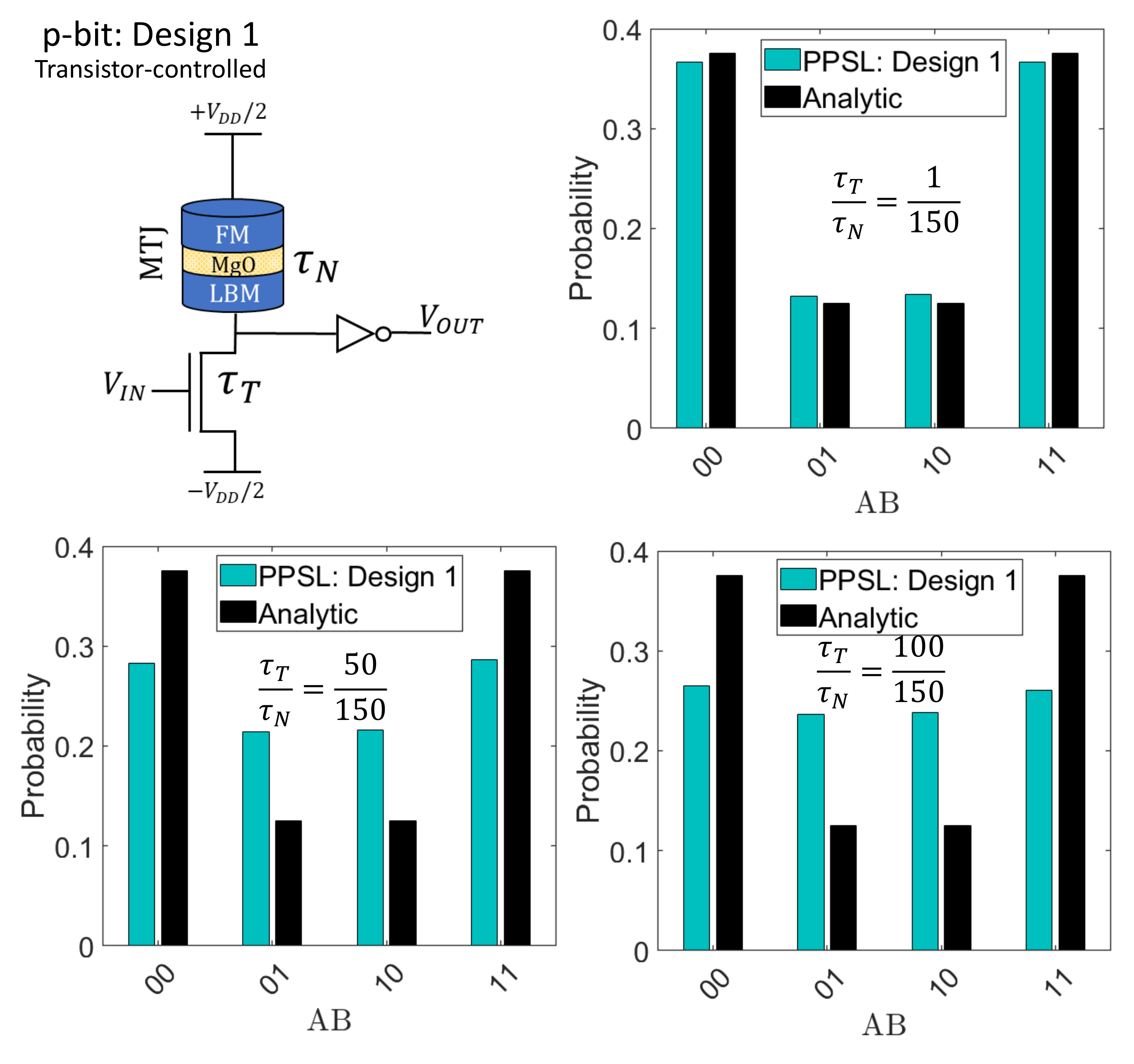}
\caption{\textbf{Effect of step response time in design 1:}The reason for design 1 to work accurately as an anutonomous Bayesian network as shown in fig.~\ref{fig2} is the two different time scales ($\tau_T$ and $\tau_N$) in this design with the condition that $\tau_T \ll \tau_N$. The same histogram shown in fig.~\ref{fig2} is plotted using the proposed behavioral model for different $\tau_T/\tau_N$ ratios and compared with the analytical values. It can be seen that as $\tau_T$ gets comparable to $\tau_N$, the probility distribution diverges from the standard statistical values.}
\label{fig3}
\end{figure}

\begin{itemize}
\item  Pinning of the s-MTJ fluctuation due to spin transfer torque (STT) effect is ignored by assuming $I_{MTJ}=0$ in eqn.~\ref{eq:Imtj}. This is a reasonable assumption considering circular in-plane magnets that are very difficult to pin due to the large demagnetization field that is always present, irrespective of the energy barrier \cite{hassan2019low}. This effect is more prominent in perpendicular anisotropy magnets (PMA) magnets. It is important to include the pinning effect in p-bits with bipolar LBM fluctuations since in this case the p-bit does not provide a sigmoidal response without the pinning current. This effect is also experimentally observed in \cite{borders2019integer} for PMA magnets. Such a p-bit design with bipolar PMA and STT pinning might not work for Bayesian networks in general, since in this case $\tau_{step}$ will be dependent on magnet fluctuation time $\tau_N$.
\item In the proposed behavioral model, the step response time of the NMOS transistor $\tau_T$ in design 1 is assumed to be independent of the input $I$. But there is a functional dependence of $\tau_T$ on $I$ in real hardware.
\item The NMOS transistor resistance $r_T$ is approximated as a $\mathrm{tanh}$ function for simplicity. In order to capture the hardware behavior in a better way, the $\mathrm{tanh}$ can be replaced by a more complicated function and the weight matrix $[J]$ will have to be learnt around that function.
\end{itemize}
All the non-ideal effects listed above are supposed to have minimal effects on different probability distributions shown in this article. Real LBMs may suffer from common fabrication defects, resulting in variations in average magnet fluctuation time $\tau_N$ \cite{abeed2019low}.The autonomous BN is also quite tolerant to such variations in $\tau_N$ as long as $\tau_T \ll min(\tau_N)$.

It is important to note that, for design 1 (Transistor-controlled) to function as a p-bit that has a step response time ($\tau_{step}$) much smaller than its average fluctuation time ($\tau_N$), the LBM fluctuation needs to be continuous and not bipolar. It is important to note that while most experimental implementations of low barrier magnetic tunnel junctions or spin-valves exhibit telegraphic (binary) fluctuations \cite{pufall2004large, locatelli2014noise, parks2018superparamagnetic, debashis2020correlated}, theoretical results \cite{kaiser2019subnanosecond, hassan2019low, abeed2019low} indicate that it should be possible to design low barrier magnets with continuous fluctuations. Preliminary experimental results for such circular disk nanomagnets have been presented in \cite{debashis2016experimental}. We believe that a lack of experimental literature on such magnets is partly due to the lack of interest of randomly fluctuating magnets that have long been discarded as impractical and irrelevant. The other experimentally demonstrated p-bits \cite{debashis2020spintronic, ostwal2019spin, ostwal2018spin} fall under design 2 category with the LBM magnetization tuned by SOT effect and are not suitable for autonomous Bayesian network operation in general. It might also be possible to design p-bits using other phenomena such as voltage controlled magnetic anisotropy (VCMA) \cite{amiri2012voltage}, but this is beyond the scope of the present study. Here we have specifically focused on two designs that can be implemented with existing MRAM technology based on STT and SOT.


\section*{Acknowledgments}

This work was supported in part by ASCENT, one of six centers in JUMP, a Semiconductor Research Corporation (SRC) program sponsored by DARPA.

\bibliographystyle{IEEEtran}
\bibliography{ABN}

\pagebreak

\widetext
\begin{center}
\textbf{\large Supplemental Materials: Hardware Design for Autonomous Bayesian Networks}
\end{center}
\setcounter{equation}{0}
\setcounter{figure}{0}
\setcounter{table}{0}
\setcounter{page}{1}
\makeatletter
\renewcommand{\theequation}{S\arabic{equation}}
\renewcommand{\thefigure}{S\arabic{figure}}
\renewcommand{\bibnumfmt}[1]{[S#1]}
\renewcommand{\citenumfont}[1]{S#1}

\section*{Additional Bayesian network example}
\begin{figure}[htbp]
\begin{center}
\includegraphics[width=0.8\linewidth]{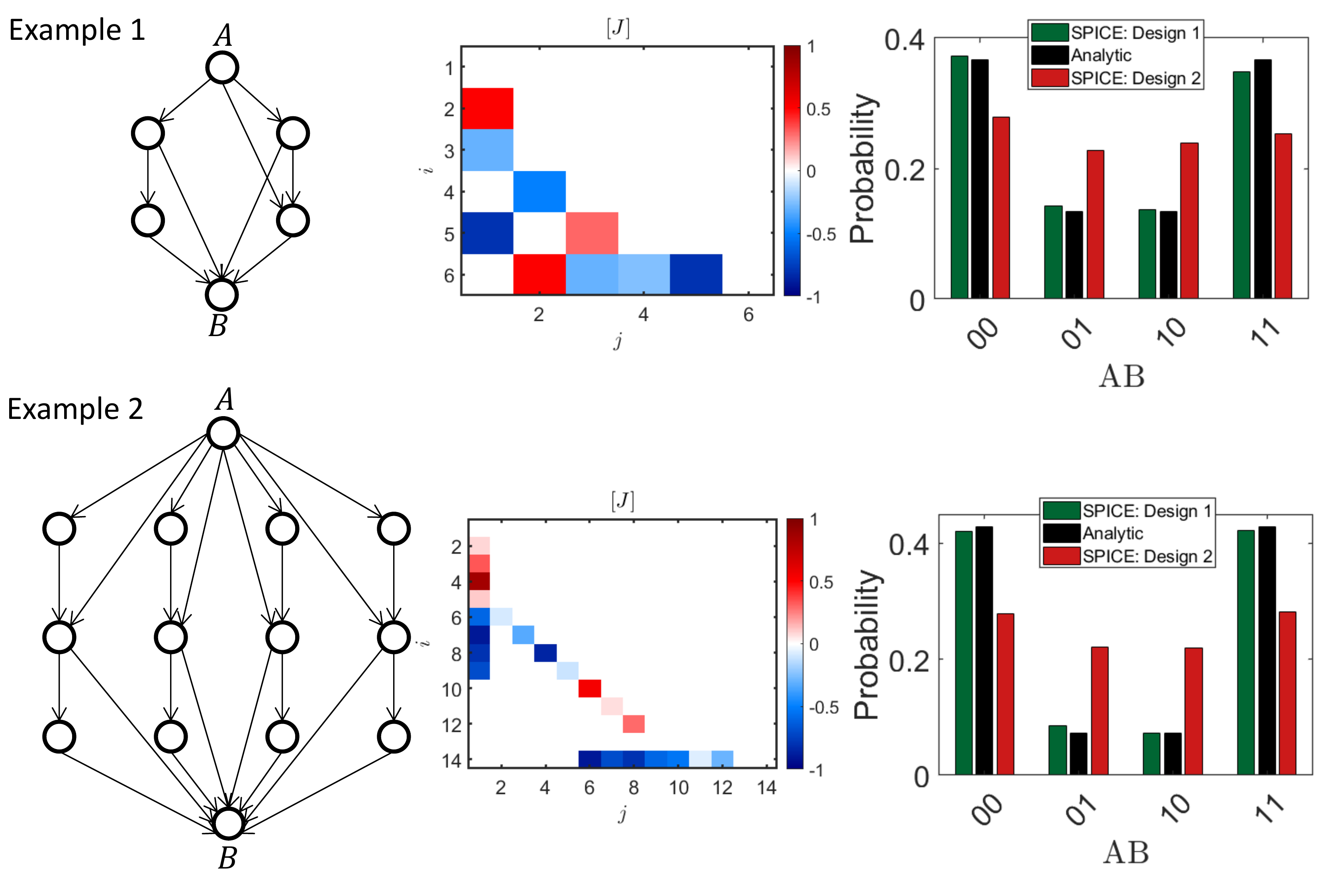}
\end{center}
\caption{\textbf{Difference between design 1 and design 2 for two different Bayesian network examples:} Example 1 is a six node network with random interconnection strength between $+1$ and $-1$ as shown in the heatmap of the coupling matrix $[J]$. Example 2 is a larger network consisting of 14 nodes with random interconnection strength between $+1$ and $-1$. Both the networks have connections not only between two consecutive layers as in common multilayer perceptron type neural networks, but also connection from other layers. SPICE simulation of both the networks implemented on an autonomous hardware with design 1 and design 2 p-bits shows that probability distribution of four configurations of nodes $(A,B)$ from design 1 matches the standard analytical results, but results from design 2 does not.}
\label{figS1}
\end{figure}

The basic conclusion presented in fig. 3 of the main manuscript is tested for other networks as well with more complex topology. Figure ~\ref{figS1} shows two more examples of Bayesian networks implemented on an autonomous hardware using two p-bit designs (design 1 and design 2) as shown in fig. 1 of the main manuscript. For both the examples, the probability distribution of four configurations of nodes $A$ and $B$ are shown in a histogram and compared with standard analytical results from applying probability chain rule. It is shown that results from design 1 autonomous hardware match well with the analytical results, but design 2 does not match. These two examples again varify the fact that design 1 autonomous hardware works for Bayesian networks in general, but design 2 does not.

\end{document}